\documentclass[a4paper,11pt]{article}
\usepackage{amsmath, amssymb, graphics, graphicx}
\usepackage{color}
\usepackage{longtable}
\usepackage{subfigure}
\usepackage{multirow}
\usepackage{amsthm}
\usepackage{bbold}
\usepackage{tcolorbox}
\usepackage{tikz}
\usetikzlibrary{shapes.geometric,shapes.multipart,arrows.meta}

\tikzstyle{box} = [rectangle, thick, rounded corners, minimum width=4.5cm, minimum height=1.5cm, text centered, draw=black]
\newcommand{\arrowIn}{\tikz \draw[-{Stealth[length=3mm,width=2mm]}] (-1pt,0) -- (1pt,0);}
\newcommand{\arrowDown}{\tikz \draw[-{Stealth[length=3mm,width=2mm]}] (0,1pt) -- (0,-1pt);}

\newcommand{\mathsym}[1]{{}}

\newcommand{\rmd}{\mathrm{d}}

\begin{document}

\begin{titlepage}       \vspace{10pt} \hfill 

\vspace{20mm}

\begin{center}

{\large \bf The Third way to 3D Supersymmetric Massive Yang-Mills Theory}

\vspace{30pt}

Nihat Sadik Deger$\,{}^{a}$ and Jan Rosseel$\,{}^{b}$
\\[6mm]

{\small
{\it ${}^a$Department of Mathematics, Bogazici University,\\ 
Bebek, 34342, Istanbul, Turkey}\\ [2mm]
{\it ${}^b$Faculty of Physics, University of Vienna,\\
Boltzmanngasse 5, A-1090, Vienna, Austria}
}

\vspace{20pt}

\end{center}

\centerline{{\bf{Abstract}}}
\vspace*{5mm}
\noindent
We construct three-dimensional, $\mathcal{N}=1$ off-shell supersymmetric massive Yang-Mills (YM) theory whose
YM equation is "third way" consistent. This means that the field equations of this model
do not come from variation of a local action without additional fields, yet the gauge covariant divergence of the YM
equation still vanishes on-shell. To achieve this, we modify the massive Majorana spinor equation so that
its supersymmetry variation gives the modified YM equation whose bosonic part coincides with the third way consistent 
YM model constructed earlier in \cite{Arvanitakis:2015oga}.

\vspace{15pt}

\end{titlepage}

\section{Introduction}

"Third way consistency" is a novel mechanism discovered in the context of three-dimensional (3D) pure gravity
theories in \cite{Bergshoeff:2014pca} where Einstein's field equation was modified with an interaction 
term that does not come from variation of a local action for the metric. Nevertheless, the modified field equation still makes
sense since the divergence of this gravitational source term vanishes if one uses the field equation again, see \cite{Bergshoeff:2015zga} for a review. Additional such 3D gravity models were obtained in \cite{Ozkan:2018cxj, Afshar:2019npk}. Meanwhile, a 3D gauge theory example was found in \cite{Arvanitakis:2015oga}. Interacting $p$-form theories with this property in general dimensions were constructed in 
\cite{Broccoli:2021pvv}. An open question regarding this class of models is whether it is possible to incorporate supersymmetry in them. In this paper we answer this question in the affirmative by explicitly constructing a supersymmetric
version of \cite{Arvanitakis:2015oga}.

The third way consistent field equation of \cite{Arvanitakis:2015oga} can be viewed as a deformation of the well-known 3D topologically massive Yang-Mills (TMYM) theory. TMYM itself was constructed a long time ago \cite{Schonfeld:1980kb, Deser:1982vy, Deser:1981wh} and its spin-2 counterparts led to many important developments in the area of 3D massive gravity theories. The model of \cite{Arvanitakis:2015oga} is then obtained by adding an extra term to the field equation of TMYM. This additional term is quadratic in the Yang-Mills (YM) field strength and can not be derived from an action with the YM field alone. While supersymmetric generalizations of TMYM have been formulated, see for example \cite{Aragone:1983xa, Zupnik:1988ry}, no supersymmetric version of its third way consistent deformation has been constructed yet. The purpose of our paper is to fill this gap, i.e., to see whether the third way and supersymmetric extensions of TMYM can be combined and whether theories in the lower-right corner of the following diagram exist:

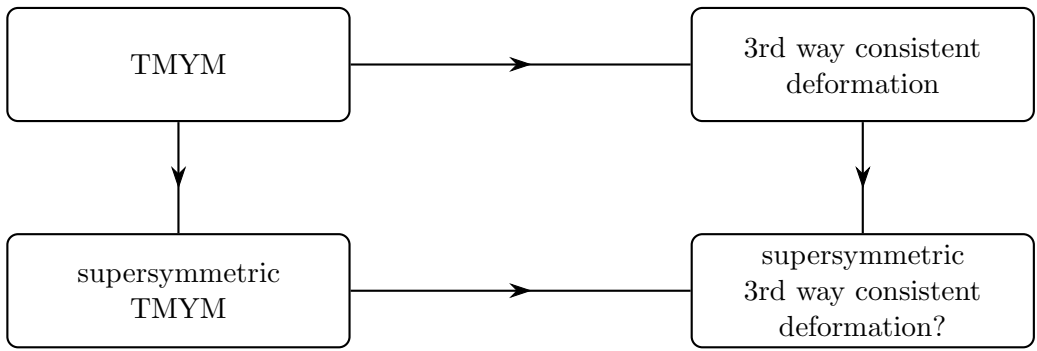
\begin{figure}[h]
\centering
\begin{tikzpicture}[node distance=9cm, text centered]
  \node (tmym) [box] {TMYM};
  \node (thirdway) [box, right of=tmym,text width=4cm] {3rd way consistent \\ deformation};
  \node (susytmym) [box, below of=tmym,text width=4cm, yshift=6cm] {supersymmetric TMYM};
  \node (this) [box, below of=thirdway, text width=4cm, yshift=6cm] {supersymmetric \\ 3rd way consistent \\ deformation?};
  \draw[thick] (tmym) -- (thirdway) node[pos=0.5]{\arrowIn};
  \draw[thick] (susytmym) -- (this) node[pos=0.5]{\arrowIn};
  \draw[thick] (tmym) -- (susytmym) node[pos=0.5]{\arrowDown};
  \draw[thick] (thirdway) -- (this) node[pos=0.5]{\arrowDown};
\end{tikzpicture}
\end{figure}

Due to the lack of an action without auxiliary fields for \cite{Arvanitakis:2015oga}, we will work with equations of motion. We will in particular attempt to construct a deformation of the equations of motion of supersymmetric TMYM, such that the resulting bosonic and fermionic equations are mapped to each other under supersymmetry. It will then be most convenient for us to work with an off-shell supersymmetric YM multiplet, for which the supersymmetry algebra closes without using the field equations. In this way we avoid having to consider modifications of the supersymmetry transformation rules of the YM multiplet. In this paper, we will study the simplest possible choice, namely a single off-shell $\mathcal{N}=1$ supersymmetric YM multiplet which only has a Majorana spinor in addition to the YM gauge field. 

We start with a brief description of the $\mathcal{N}=1$ off-shell supersymmetric TMYM
in the next section. In section 3, we first review the
model of \cite{Arvanitakis:2015oga}. Then, we make a modification of the spinor 
field equation of the  $\mathcal{N}=1$ supersymmetric TMYM theory so that its supersymmetry
variation gives a modified YM-equation whose bosonic part is the model found in \cite{Arvanitakis:2015oga}.
The full description of our model is given in equation \eqref{eq:summaryeoms}. It has one extra mass parameter compared
to TMYM \cite{Schonfeld:1980kb, Deser:1982vy, Deser:1981wh} like \cite{Arvanitakis:2015oga} and when fermions
are set to zero it reduces to \cite{Arvanitakis:2015oga}. We finish this section by showing that our modified YM-equation
is third way consistent. We conclude with some comments in section 4. 

\section{A Review of 3D, $\mathcal{N}=1$ Topologically Massive super-Yang-Mills}

In this paper we would like to construct the $\mathcal{N}=1$ off-shell supersymmetric version of the third way consistent massive 3D YM theory of \cite{Arvanitakis:2015oga}. Since this theory will correspond to a deformation of the $\mathcal{N}=1$ supersymmetric TMYM theory, we will first review the latter here. This section also serves to introduce the notation and conventions used in the rest of this letter.

We will consider YM theory for an arbitrary non-abelian gauge group $G$ with structure constants $f^I_{JK}$ (with $I$, $J$, $K = 1,\cdots, \mathrm{dim}(G)$). The off-shell $\mathcal{N}=1$ super-YM multiplet then consists of the gauge field $A_\mu^I$ and a Majorana spinor $\chi^I$, both transforming in the adjoint of $G$. In our conventions, a gauge transformation with parameter $\Lambda^I$ of these fields is given by
\begin{equation}
  \label{eq:gaugetrafos}
  \delta A_\mu^I = \partial_\mu \Lambda^I - f^I_{JK} \Lambda^J A_\mu^K \,, \qquad \qquad \qquad \delta \chi^I = - f^I_{JK} \Lambda^J \chi^K \,.
\end{equation}
As a consequence, the gauge covariant field strength $F^I_{\mu\nu}$ and covariant derivative $D_\mu X^I$ of any object $X^I$ in the adjoint representation of $G$ are explicitly given by
\begin{equation}
  \label{eq:covquantities}
  F^I_{\mu\nu} = 2 \partial_{[\mu} A_{\nu]}^I + f^I_{JK} A^J_\mu A^K_\nu \,, \qquad \qquad \qquad  D_\mu X^I = \partial_\mu X^I + f^I_{JK} A_\mu^J X^K \,.
\end{equation}
Note that the following Bianchi identity then holds: $D_{[\mu} F^I_{\nu\rho]} = 0 \,.$

The supersymmetry transformation rules of $A^I_\mu$ and $\chi^I$ are given by\footnote{Our conventions are as follows:
The charge conjugation matrix $C$ obeys $\gamma^T_\mu = -C \gamma_\mu C^{-1} \,, \quad C^T = -C \,.$
The following properties for bilinears involving Majorana spinors then hold:
$\bar{\chi} \epsilon = \bar{\epsilon} \chi \,, \quad \bar{\chi} \gamma^\mu \epsilon = - \bar{\epsilon} \gamma^\mu \chi \,.$ The Fierz identity reads:
$\chi \bar{\epsilon} = -\frac{1}{2} \bar{\epsilon}\chi \mathbb{1} - \frac{1}{2} \bar{\epsilon} \gamma^\mu \chi \gamma_\mu \,.$
Some useful $\gamma$-matrix relations are:
$\gamma^\nu \gamma^\mu \gamma_\nu = - \gamma^\mu ,\quad \gamma^{\mu\nu} = \epsilon^{\mu\nu\rho} \gamma_\rho ,  \quad \gamma_\mu = -\frac{1}{2} \epsilon_{\mu\nu\rho} \gamma^{\nu\rho} , \quad \gamma^{\mu\nu\rho} = \epsilon^{\mu\nu\rho} \mathbb{1} \,.$}
(see e.g. \cite{Andringa:2009yc})
\begin{equation}
  \label{eq:susytrafos}
  \delta A_\mu^I = - \bar{\epsilon} \gamma_\mu \chi^I \,, \qquad \qquad \qquad \delta \chi^I = \frac18 \gamma^{\mu\nu} F^I_{\mu\nu} \epsilon \,,
\end{equation}
where the Majorana spinor $\epsilon$ denotes the supersymmetry parameter. One can check that the supersymmetry algebra then indeed closes off-shell on the fields $A^I_\mu$ and $\chi^I$.

The equations of motion of the 3D, $\mathcal{N}=1$ Topologically Massive super-YM theory are derived from the following action
\begin{align}
  \label{eq:TMSYM}
  S = \frac{1}{g^2} \int \rmd^3 x  \Big( -\frac14 F^I_{\mu\nu} F^{\mu\nu}_I - 2 \bar{\chi}^I \gamma^\mu D_\mu \chi_I + \mu \left[ \epsilon^{\mu\nu\rho} A^I_\mu \left( \partial_\nu A_{\rho\, I} + \frac13 f_{KLI} A^K_\nu A^L_\rho \right) 
  - 4 \bar{\chi}^I \chi_I \right] \Big) .
\end{align}
Here, $g$ denotes the YM coupling constant, $\mu$ is a mass parameter and we have lowered the adjoint index $I$, using the invariant Cartan-Killing metric $\eta_{IJ}$ (via e.g. $\chi_I = \chi^J \eta_{JI}$ and $f_{KLI} = f^J_{KL} \eta_{JI}$). Note furthermore that $g^2$, $\mu$ and $A_\mu^I$ have mass dimension 1, while $\chi^I$ has mass dimension $\frac32$. When setting the fermion $\chi^I$ to zero, the action \eqref{eq:TMSYM} describes the TMYM theory considered in \cite{Schonfeld:1980kb, Deser:1982vy, Deser:1981wh}.

One can check that \eqref{eq:TMSYM} is invariant under the supersymmetry transformation rules \eqref{eq:susytrafos}. The equations of motion that follow from the action \eqref{eq:TMSYM} are given by
\begin{align}
  & D_\nu F^{\nu\mu\, I} + \mu\, \epsilon^{\mu\nu\rho} F^I_{\nu\rho} = - 2 f^I_{KL} \bar{\chi}^K \gamma^\mu \chi^L \,, \label{eq:boseomTMSYM} \\
  & \gamma^\mu D_\mu \chi^I + 2 \mu\, \chi^I = 0 \,. \label{eq:fermeomTMSYM}
\end{align}
Note that, when contracted with $D_\mu$ both terms on the left-hand-side of \eqref{eq:boseomTMSYM} give identically zero. For consistency, the covariant divergence of the source current
\begin{align}
  \label{eq:defcurrent}
  j_\mu^I \equiv f^I_{JK} \bar{\chi}^J \gamma_\mu \chi^K
\end{align}
on the right-hand-side of \eqref{eq:boseomTMSYM} should then also vanish and it is easy to see that this is true on-shell.

For what follows, we note that the supersymmetry transformation of the fermionic equation of motion \eqref{eq:fermeomTMSYM} leads to the bosonic one \eqref{eq:boseomTMSYM}, as follows:
\begin{align} \label{eq:varfermeomTMSYM}
  \delta \left(\gamma^\mu D_\mu \chi^I + 2 \mu \chi^I \right) &= \frac14 \Big( D^\nu F_{\nu\mu}^I + \mu\, \epsilon_{\mu}{}^{\nu\rho} F^I_{\nu\rho} + 2 f^I_{JK} \bar{\chi}^J \gamma_\mu \chi^K \Big) \gamma^\mu \epsilon \,.
\end{align}
The fact that the supersymmetry variation of the spinor field equation is of the form $Y_\mu \gamma^\mu \epsilon$
where $Y_\mu=0$ is the bosonic field equation, will be our guiding principle in constructing the supersymmetric version
of the model of \cite{Arvanitakis:2015oga} in the next section.

\section{Third Way Consistent 3D, $\mathcal{N}=1$ Massive super-Yang-Mills}

We first briefly review the third way consistent massive YM theory constructed in \cite{Arvanitakis:2015oga}.
Its source free field equation for an arbitrary gauge group $G$ is obtained by adding an extra term to the equation of motion of TMYM (itself obtained by setting $\chi^I$ to zero in \eqref{eq:boseomTMSYM}):
\begin{align}
\label{MYM}
\epsilon_\mu{}^{\nu\rho} D_\nu \tilde{F}^I_\rho + 2 \mu \tilde{F}^I_\mu + \frac{2}{m} \epsilon_\mu{}^{\nu\rho} f^I_{JK} \tilde{F}^J_\nu \tilde{F}^K_\rho = 0 \,, 
\end{align}
where, $m$ is another mass parameter and we introduced the dual field strength $\tilde{F}_\mu^I$ notation as
\begin{align}
  \label{eq:defdualF}
  \tilde{F}_\mu^I = \frac12 \epsilon_\mu{}^{\nu\rho} F^I_{\nu \rho} \qquad \Leftrightarrow \qquad F^I_{\mu\nu} = -\epsilon_{\mu\nu}{}^\rho \tilde{F}^I_{\rho} \,.
\end{align}
It is easy to see that this equation can not be the Euler-Lagrange equation of a gauge-invariant local action
for the YM vector field alone \cite{Arvanitakis:2015oga} and hence its consistency is not automatic. Indeed, 
if we hit \eqref{MYM} with the covariant derivative $D^\mu$ the first two terms are identically zero whereas the interaction term
is not. However, if we use \eqref{MYM} again it vanishes due to the Jacobi identity, which is the essence of the third way consistency mechanism. The special cases $\mu=0$ and $\mu=2m$ of this model were studied earlier in \cite{Mukhi:2011jp} and \cite{Nilsson} respectively.  

It is possible to add a matter current $\mathcal{J}_\mu^I$ to \eqref{MYM} as
\begin{align}
\label{eq:genboseom}
\epsilon_\mu{}^{\nu\rho} D_\nu \tilde{F}^I_\rho + 2 \mu \tilde{F}^I_\mu + \frac{2}{m} \epsilon_\mu{}^{\nu\rho} f^I_{JK} \tilde{F}^J_\nu \tilde{F}^K_\rho = \mathcal{J}_\mu^I \,, 
\end{align}
provided that it satisfies
\begin{align} \label{eq:thirdwayconsistency}
  D_\mu \mathcal{J}^{\mu\, I} + \frac{4}{m} f^I_{JK} \tilde{F}_\mu^J \mathcal{J}^{\mu\, K} = 0 \,,
\end{align}
to maintain the third way consistency of the model \cite{Arvanitakis:2015oga}.

In this paper our aim is to obtain a $\mathcal{N}=1$ off-shell supersymmetric extension of \eqref{MYM}. Since
\eqref{MYM} does not come from an action without extra auxiliary fields, we will do this using the Noether procedure
at the level of equations of motion. The advantage of having off-shell
supersymmetry is that the supersymmetry variations \eqref{eq:susytrafos} remain the same for any extension of 
the field equations \eqref{eq:boseomTMSYM} and \eqref{eq:fermeomTMSYM}. Notice that \eqref{MYM} is simply a deformation of
\eqref{eq:boseomTMSYM} when fermions are set to zero. Therefore, to achieve our goal
we will start from the fermionic equation of motion \eqref{eq:fermeomTMSYM} of the $\mathcal{N}=1$ supersymmetric TMYM theory
and modify it so that its supersymmetry variation is of the form $Y_\mu \gamma^\mu \epsilon$ 
with the bosonic part of $Y_\mu=0$ given by \eqref{MYM}. The full $Y_\mu=0$ equation will be the new YM equation 
as happened in \eqref{eq:varfermeomTMSYM} and this together with the modified 
spinor equation will define our supersymmetric model. In this case, the $Y_\mu=0$ equation will be of the form 
\eqref{eq:genboseom} and therefore the fermionic current $\mathcal{J}_\mu^I$ that we will find should satisfy the third way consistency condition
\eqref{eq:thirdwayconsistency}. Moreover, the supersymmetry variation of the $Y_\mu=0$ equation should give the derivative of the
new spinor equation on-shell.

Note that \eqref{MYM} contains  $\epsilon_\mu{}^{\nu\rho} f^I_{JK} \tilde{F}^J_\nu \tilde{F}^K_\rho$ as an additional
term to \eqref{eq:boseomTMSYM} and the supersymmetry variation of the spinor \eqref{eq:susytrafos} immediately
suggests adding a term of the form $f^I_{JK} \tilde{F}^J_{\mu} \gamma^{\mu} \chi^K$ to the spinor field equation 
\eqref{eq:fermeomTMSYM}. However, the supersymmetry variation of this term gives, in addition to what we want,
a term of the form $f^I_{JK}\bar{\chi}^J \gamma^\mu D_\mu \chi^K$.  Due to the modification of the spinor field equation
this term is neither zero on-shell, nor of the form that we want. Hence, we need to modify the spinor field equation
further to cancel this term, either identically or on-shell. The form of this unwanted piece indicates that on-shell cancellation can be achieved by adding an extra term that is cubic in $\chi^I$.
Therefore, we consider the following modification of \eqref{eq:fermeomTMSYM} as a candidate equation of motion of $\chi^I$:
\begin{align} \label{eq:fermeomMMSYM}
  \Psi^I \equiv \gamma^\mu D_\mu \chi^I + 2 \mu \chi^I + \frac{2}{m} f^I_{JK} \tilde{F}^J_{\mu} \gamma^{\mu} \chi^K
  + a f^I_{JK} f^K_{MN} \bar{\chi}^M \gamma^\mu \chi^N \gamma_\mu \chi^J =0 \,,
\end{align}
where $``a"$ is a constant to be determined. As mentioned above, we determine $a$ by requiring that the supersymmetry transformation of $\Psi^I$ is on-shell of the form $Y_\mu \gamma^\mu \epsilon$. After repeatedly using Fierz, Jacobi and $\gamma$-matrix identities, we find that with the choice $a= \frac{16}{3m^2}$ one gets
\begin{align}\label{spinorsusy}
  \delta \Psi^I  = \frac14 \Xi^I_\mu \gamma^\mu \epsilon - \frac{4}{m} f^I_{JK} \bar{\chi}^J \epsilon \Psi^K  \, ,
\end{align}
where
\begin{align} \label{eq:boseomMMSYM2}
  &\Xi^I_\mu \equiv D^\nu F_{\nu\mu}^I + \mu \epsilon_\mu{}^{\nu\rho} F_{\nu\rho}^I - \frac{1}{m} f^I_{JK} \epsilon^{\rho \sigma \nu} F^J_{\nu\mu} F^K_{\rho\sigma} + \left(2 - \frac{16\mu}{m}\right) f^I_{JK} \bar{\chi}^J \gamma_\mu \chi^K + \frac{8}{m} \epsilon_{\mu}{}^{\nu\rho} f^I_{JK} \bar{\chi}^J \gamma_\nu D_\rho \chi^K \nonumber \\ & \qquad \qquad + \frac{16}{m^2} f^I_{JK} f^K_{MN} \bar{\chi}^M \gamma^\nu \chi^N F_{\nu\mu}^J  + \frac{32}{m^3} \epsilon_\mu{}^{\nu\rho} f^I_{KL} f^K_{JO} f^L_{MN} \bar{\chi}^J \gamma_{\nu} \chi^O \bar{\chi}^M \gamma_\rho \chi^N  \, .
\end{align}
A lengthy computation then shows that the supersymmetry variation of $\Xi^I_\mu$ is given by
\begin{align} \label{xisusy}
  \delta \Xi^I_\mu &= -\bar{\epsilon} \gamma_\mu{}^\nu D_\nu \Psi^I + \frac{4}{m} f^I_{JK} \bar{\epsilon} \chi^K \Xi^J_\mu  - \frac{16}{m^2} f^I_{JK} f^K_{MN} \bar{\chi}^M \gamma^\nu \chi^N \bar{\epsilon} \gamma_{\mu\nu} \Psi^J \,.
\end{align}
Together with \eqref{spinorsusy}, we thus see that $\Psi^I$ and $\Xi^I_\mu$ transform into each other under supersymmetry. We can then propose $\Psi^I = 0$ and $\Xi^I_\mu = 0$ as a supersymmetric set of equations of motion. Since the bosonic part of $\Xi^I_\mu = 0$, i.e. the first three terms in \eqref{eq:boseomMMSYM2}, coincides with that of the equation of motion of pure massive Yang-Mills theory \cite{Arvanitakis:2015oga} given in \eqref{MYM}, we see that $\Psi^I = 0$ and $\Xi^I_\mu = 0$ can be identified as the equations of motion of the $\mathcal{N}=1$ supersymmetric version of \eqref{MYM}. Note however that in the presence of supersymmetry, the equation of motion \eqref{MYM} of the pure theory gets modified by extra terms that represent a coupling of the spin-1 gauge vector $A_\mu^I$ to the spin-1/2 gaugino $\chi^I$.

Summarizing, we propose the following equations of motion for the third way consistent 3D, $\mathcal{N}=1$  massive super-Yang-Mills theory:
\begin{align}
  \label{eq:summaryeoms}
  & \epsilon_\mu{}^{\nu\rho} D_\nu \tilde{F}^I_\rho + 2 \mu \tilde{F}^I_\mu + \frac{2}{m} \epsilon_\mu{}^{\nu\rho} f^I_{JK} \tilde{F}^J_\nu \tilde{F}^K_\rho = \mathcal{J}_\mu^I \,, \nonumber \\
  & \gamma^\mu D_\mu \chi^I + 2 \mu \chi^I + \frac{2}{m} f^I_{JK} \tilde{F}^J_{\mu} \gamma^{\mu} \chi^K + \frac{16}{3m^2} f^I_{JK} f^K_{MN} \bar{\chi}^M \gamma^\mu \chi^N \gamma_\mu \chi^J = 0 \,, \nonumber \\ 
  & \quad \mathrm{with}\ \ \ \mathcal{J}_\mu^I = \left(\frac{16\mu}{m} - 2 \right) j_\mu^I + \frac{4}{m} \epsilon_\mu{}^{\nu\rho} D_\nu j_\rho^I + \frac{16}{m^2} \epsilon_\mu{}^{\nu\rho} f^I_{JK} \tilde{F}_\nu^J j_\rho^K - \frac{32}{m^3} \epsilon_\mu{}^{\nu\rho} f^I_{JK} j_\nu^J j_\rho^K \,, \nonumber \\
  & \quad \mathrm{where}\ \ 
  j_\mu^I = f^I_{JK} \bar{\chi}^J \gamma_\mu \chi^K \,.
\end{align}

We still need to verify that our YM equation is indeed third way consistent. This will be done in the next subsection.

\subsection{The `Third way' Consistency}

Note that our matter source $\mathcal{J}_\mu^I$ that appears in \eqref{eq:summaryeoms} is of the form
\begin{align} \label{eq:genJ}
  \mathcal{J}_\mu^I &= c_1 j_\mu^I + c_2 \epsilon_\mu{}^{\nu\rho} D_\nu j_\rho^I + c_3 \epsilon_\mu{}^{\nu\rho} f^I_{JK} \tilde{F}_\nu^J j_\rho^K + c_4 \epsilon_\mu{}^{\nu\rho} f^I_{JK} j_\nu^J j_\rho^K \,,
\end{align}
where $c_1 =  \frac{16\mu}{m} - 2 \,,  c_2 = \frac{4}{m} \,, c_3 = \frac{16}{m^2} \,,  c_4 = -\frac{32}{m^3} \,.$
To have a third way consistent system, $\mathcal{J}_\mu^I$
should satisfy \eqref{eq:thirdwayconsistency} as
we discussed above. The form of the current $\mathcal{J}_\mu^I$ in terms of $j_\mu^I$ 
is the same as in \cite{Arvanitakis:2015oga}, however unlike \cite{Arvanitakis:2015oga} we will not assume that $j_\mu^I$ is conserved. Using \eqref{eq:genJ}, one can derive that on-shell the following holds
\begin{align} \label{eq:thirdwayconsistency1}
  D_\mu \mathcal{J}^{\mu\, I} + \frac{4}{m} f^I_{JK} \tilde{F}^{\mu\, J} \mathcal{J}_\mu^K &= c_1 D_\mu j^{\mu\, I} + \left(c_2 - 2 \mu c_3  + \frac{4}{m}c_1 \right) f^I_{JK} \tilde{F}^J_\mu j^{\mu\, K} \nonumber
   \\ & \quad + \left(2 c_4 + c_2 c_3\right) \epsilon^{\mu\nu\rho} f^I_{JK} D_\mu j_\nu^J j_\rho^K + \left(c^2_3 + \frac{8}{m}c_4 \right) \epsilon^{\mu\nu\rho} f^I_{JK} f^J_{MN} \tilde{F}^M_\mu j_\nu^N j_\rho^K \nonumber \\ & \quad + \left(c_3 -  \frac{4}{m}c_2  \right) \epsilon^{\mu\nu\rho} f^I_{JK} D_\mu j_\nu^J \tilde{F}^K_\rho \,.
\end{align}
Note that in order to derive this result, we needed to use the bosonic equation of motion \eqref{eq:genboseom}.
For our $j_\mu^I$ given in \eqref{eq:defcurrent} $D_\mu j^{\mu\, I} \neq 0$, instead, an explicit computation (using the fermionic equation of motion of \eqref{eq:summaryeoms}) gives that
\begin{align}
  D_\mu j^{\mu\, I} &= - \frac{2}{m} f^I_{JK} \tilde{F}^{\mu\, J} j_\mu^K + 2 f^I_{JK} \bar{\chi}^J \Psi^K \,.
\label{option3}
\end{align}
This shows that we can replace $D_\mu j^{\mu\, I}$ on-shell by $-\frac{2}{m} f^I_{JK} \tilde{F}^{\mu\, J} j_\mu^K$. Using this in \eqref{eq:thirdwayconsistency1}, it is easy to see that coefficients of all terms in the right-hand-side of \eqref{eq:thirdwayconsistency1} vanish and
hence the consistency condition \eqref{eq:thirdwayconsistency} is satisfied for our model.

It was realized in \cite{Broccoli:2021pvv} that the model constructed in \cite{Arvanitakis:2015oga} can be found
starting from the flat connection equation $F^I_{\mu\nu}=0$ and then shifting the connection $A^M_\mu$ with an arbitrary 
linear combination of $\tilde{F}^M_\mu$ and $j^M_\mu$. After this, it is possible to add a multiple of $\tilde{F}^M_\mu$ and $j^M_\mu$ to this equation provided
that either $D^\mu j^M_\mu=0$ as in \cite{Arvanitakis:2015oga}
or  $D^\mu j^M_\mu$ is on-shell proportional to one of the terms on the right hand side of \eqref{eq:thirdwayconsistency1}
as in \eqref{option3}
without spoiling the third way consistency. This also shows that $\mathcal{J}_\mu^I$ will always have the structure
given in \eqref{eq:genJ} in terms of $j_\mu^I$. It is remarkable that this mechanism works for the spinor equation in
our model as well. Indeed
a comparison of the spinor field equation of the Topologically Massive super-Yang-Mills theory \eqref{eq:fermeomTMSYM} with
ours \eqref{eq:fermeomMMSYM} shows that the latter can be obtained from the former by shifting $A^M_\mu$ as
\begin{align}\label{shift}
A^M_\mu \rightarrow A^M_\mu +  \alpha \tilde{F}^M_\mu + \beta  j^M_\mu\, ,
\end{align}
where constants are fixed uniquely as $\alpha= \frac{2}{m}$ and $\beta= - \frac{16}{3m^2}$ 
by requiring the supersymmetry.

\section{Conclusion}
The theory presented in \eqref{eq:summaryeoms} is the main result of this paper which is the first example
of a supersymmetric third way consistent model. It is clear that its equations of motion can not be derived 
from a local action, with the field content considered in this letter. It is however conceivable that an action with auxiliary fields exists as in \cite{Arvanitakis:2015oga} and it
would be interesting to investigate this further. In this regard, it is useful to note that such an auxiliary field formulation likely includes a fermionic auxiliary field, as is suggested by how the fermionic equation of motion appears in its own supersymmetry transformation (see \eqref{spinorsusy}).\footnote{We are grateful to Paul Townsend for making this suggestion.} It would also be interesting to see whether a superspace formulation of the results presented here can be obtained. There are also various extensions of our model to consider such as  
coupling with $\mathcal{N}=1$ supersymmetric scalar or gravity multiplets and constructing $\mathcal{N}>1$ supersymmetric
versions. In \cite{Broccoli:2021pvv} it was shown that one can obtain higher derivative extensions of \cite{Arvanitakis:2015oga}
by shifting the connection \eqref{shift} with further terms which are not necessarily conserved. It would be interesting to see whether a supersymmetric extension would still be possible for such deformations. 

We hope that our construction will provide some insight for finding supersymmetric versions of the third way consistent 
gravity \cite{Bergshoeff:2014pca, Ozkan:2018cxj, Afshar:2019npk} and $p$-form theories \cite{Broccoli:2021pvv}. As we saw in \eqref{shift} the extra terms that appear in our model can be understood as coming from shifting the gauge connection $A_\mu^I$ with 
bosonic and fermionic current 1-forms of the initial theory. We expect this to be a key feature of all such models.
In 3D gravity examples \cite{Bergshoeff:2014pca, Ozkan:2018cxj, Afshar:2019npk} the shift occurs in the spin connection
in their first order formulation and we anticipate this to be supplemented with appropriate fermionic current terms in the
supersymmetric case. Observe that the supersymmetry variation of our spinor field equation involves not only the YM field equation but also contains a term proportional to itself (see \eqref{spinorsusy}) and a similar conclusion holds for the YM field equation (see \eqref{xisusy}). This could be a generic feature of third way consistent supersymmetric theories.

The model of \cite{Arvanitakis:2015oga} exhibits a Higgs mechanism \cite{Mukhi:2011jp, Nilsson} and is also related 
to multi M2-branes of 11D supergravity \cite{Bagger:2012jb}. Moreover, presence of higher derivative terms in the bosonic
third way consistent models can be understood as spontaneous breaking of a local symmetry
as was illustrated for \cite{Bergshoeff:2014pca} in \cite{Chernyavsky:2020fqs}. It would be nice to clarify
these connections for our supersymmetric model too.

\section*{Acknowledgements}
We would like to thank Eric Bergshoeff, Dmitri Sorokin, Stefan Theisen and Paul Townsend for useful feedbacks on the manuscript. We are 
grateful to the Erwin Schr\"odinger Institute (ESI), Vienna where this work was initiated in the framework of the `Research in Teams' Programme. NSD wishes to thank Albert-Einstein-Institute, Potsdam where
part of this work was carried out.


\end{document}